\documentclass[useAMS,usenatbib]{mn2e}
\usepackage{graphicx}
\usepackage{subfigure}
\usepackage{float}
\usepackage{color}
\usepackage{ amssymb }
\usepackage{ amsmath }
\usepackage{rotating}
\usepackage{bm} 
\usepackage{mathtools}

\title[The MRI with non-ideal effects in protoplanetary disks ]{Global multifluid simulations of the magnetorotational instability in radially stratified protoplanetary disks}
\author[Rodgers-Lee, Ray, Downes]{D. Rodgers-Lee$^{1,2}$\thanks{E-mail:
donna@cp.dias.ie}, T. Ray$^{1,2}$, T. P. Downes$^{3,4}$ \\ 
$^{1}$ School of Cosmic Physics, Dublin Institute for Advanced Studies, 31 Fitzwilliam Place, Dublin 2, Ireland \\
$^{2}$ School of Physics, Trinity College Dublin, Dublin 2, Ireland \\
$^{3}$ School of Mathematical Sciences, Dublin City University, Glasnevin, Dublin 9, Ireland\\
$^{4}$ National Centre for Plasma Science and Technology, Dublin City University, Glasnevin, Dublin 9, Ireland }
\begin{document}
\date{Accepted xxxx xxxxxx xx. Received xxxx xxxxxx xx; in original form xxxx xxx xx}

\pagerange{\pageref{firstpage}--\pageref{lastpage}} \pubyear{xxxx}

\maketitle

\label{firstpage}

\begin{abstract}
The redistribution of angular momentum is a long standing problem in our understanding of protoplanetary disk (PPD) evolution. The magnetorotational instability (MRI) is considered a likely mechanism. We present the results of a study involving multifluid global simulations including Ohmic dissipation, ambipolar diffusion and the Hall effect in a dynamic, self-consistent way. We focus on the turbulence resulting from the non-linear development of the MRI in radially stratified PPDs and compare with ideal MHD simulations.

In the multifluid simulations the disk is initially set up to transition from a weak Hall dominated regime, where the Hall effect is the dominant non-ideal effect but approximately the same as or weaker than the inductive term, to a strong Hall dominated regime, where the Hall effect dominates the inductive term. As the simulations progress a substantial portion of the disk develops into a weak Hall dominated disk. We find a transition from turbulent to laminar flow in the inner regions of the disk, but without any corresponding overall density feature.

We introduce a dimensionless parameter, $\alpha_\mathrm{RM}$, to characterise accretion with $\alpha_\mathrm{RM} \gtrsim 0.1$ corresponding to turbulent transport. We calculate the eddy turnover time, $t_\mathrm{eddy}$, and compared this with an effective recombination timescale, $t_\mathrm{rcb}$, to determine whether the presence of turbulence necessitates non-equilibrium ionisation calculations. We find that $t_\mathrm{rcb}$ is typically around three orders of magnitude smaller than $t_\mathrm{eddy}$. Also, the ionisation fraction does not vary appreciably. These two results suggest that these multifluid simulations should be comparable to single fluid non-ideal simulations.
\end{abstract}

\begin{keywords}
accretion, accretion discs -- turbulence -- instabilities -- methods:numerical -- (magnetohydrodynamics) MHD -- stars: low-mass -- protoplanetary discs

\end{keywords}

\section{Introduction}
\label{sec:intro}

Circumstellar disks are ubiquitous around young stellar objects (YSOs) and were first observed around YSOs nearly 30 years ago \citep{cohen_1985,harvey_1985}. Material is known to accrete from the circumstellar disk onto the protostar with a typical accretion rate of $10^{-7}\,M_\odot \ \mathrm{yr^{-1}}$, although rates are known to increase enormously during the so-called FUOR phase \citep[see, for example,][]{audard-2014}. One of the main puzzles since the first observations of YSOs has been to explain how material accretes through the disk and onto the young protostar while obeying conservation of angular momentum. Solving this problem is of great importance to advance our understanding of star, and planet, formation in general. While some authors have suggested winds/outflows launched from the disk may play a role \citep[e.g.][]{pudritz-2009} in redistributing angular momentum vertically above the disk, others have proposed turbulence \citep[e.g.][]{salmeron-2007} as a means of redistributing it radially.

The magnetorotational instability (MRI) enables turbulence to develop and has attracted much attention since its rediscovery in an astrophysical context \citep{balbus_1991,hawley_1991}. In the limit of ideal magnetohydrodynamics (MHD) the MRI can develop in a protoplanetary disk (PPD) provided two criteria are satisfied: first, a weak vertical magnetic field is present initially and second, angular velocity decreases as a function of radius. The second condition is satisfied if the disk is rotating with keplerian or near-keplerian velocities. It is also plausible to assume that PPDs are threaded by magnetic fields since the molecular clouds from which the disks form are observed to have magnetic fields strengths of a few $\mu$G \citep{crutcher_2012}. Therefore the MRI can feasibly facilitate the development of turbulence in PPDs which is an effective mechanism for redistributing angular momentum.

A range of local shearing-box \citep{hawley_1991,hawley_1996,brandenburg_1996,stone_1996,miller_2000} and global \citep{armitage_1998,hawley_2000,hawley_2001,steinacker_2002,papaloizou_2003,fromang_2006,flock_2011} simulations have been performed of the MRI in the limit of ideal MHD. There are still ongoing issues concerning ideal MHD simulations of the MRI with zero net flux such as numerical convergence \citep{fromang_2007,bodo-2014} and the dependence of results on the magnetic Prandtl number \citep{fromang-2010}. With these significant caveats, ideal MHD simulations have found values for the viscous stress parameter ($\alpha$, which is a measure of angular momentum transport) in the range of $\sim\,$0.001 - 0.1. These values for $\alpha$ can be compared to those inferred from observations of disks of known mass and accretion rate and are in broad agreement, though the issues mentioned undoubtedly raise questions as to the significance of the apparent agreement.

The reason for continued research in this area came from the realisation that PPDs are only weakly ionised \citep{hayashi_1981} and so the contribution from non-ideal MHD effects are non-negligible \citep{gammie_1996,stone_2000,wardle_2012}. The three relevant non-ideal effects are Ohmic dissipation, ambipolar diffusion and the Hall effect. 

Ambipolar diffusion dominates in areas of low density and is due to the poor coupling, via collisions, of charged and neutral particles leading to a drift in their relative positions perpendicular to an electric or magnetic field. When the density becomes very large, as at the disk midplane, this effect becomes isotropic and is referred to as Ohmic dissipation. The Hall effect is important at intermediate densities and arises due to a difference in velocities between electrons and ions. This differential velocity occurs because the charged species have different charge-to-mass ratios and collision coefficients meaning they are coupled by differing amounts to magnetic field lines. These non-ideal effects must be taken into account since their associated length scales are comparable to the characteristic length scales which describe PPDs, such as the disk height \citep[see review by][]{turner-2014}. 

Simulations of Ohmic dissipation \citep{jin_1996,fleming_2000,sano_2001} and ambipolar diffusion \citep{blaes_1994,maclow_1995,hawley_1998} have shown that both of these non-ideal effects diffuse the magnetic field leading to a suppression of the MRI. Such severely damped MRI driven turbulence cannot in most cases account for the observed accretion rates. 

The last non-ideal effect to be considered is the Hall effect which, unlike the other non-ideal effects, does not diffuse magnetic fields. Linear stability analysis of the MRI in the presence of the Hall effect showed that it should have either a destabilising or stabilising effect on the MRI depending on the orientation of the magnetic field ($\bold B$) with respect to the angular velocity vector ($\bold \Omega$) of the disk \citep{wardle_1999,balbus_2001}. Early results of simulations of the MRI that included the Hall effect \citep{sano_2002a,sano_2002b} did not appear to reflect this but this was later explained by \citet{wardle_2012} as being due to strong Ohmic dissipation.

Continued study of the Hall effect has evidenced a wide variety of behaviour depending on its strength relative to other non-ideal effects and the inductive term. When $\bold \Omega \cdot \bold B >0$ a strong Hall effect has been shown to lead to zonal flows \citep{kunz_2013,bai_2015,bethune-2016} with little accretion, or laminar flow with strong Maxwell stresses but significant accretion \citep{lesur_2014}. A weaker Hall effect leads to the expected canonical behaviour of fully developed MRI driven turbulence with a high accretion rate \citep{okeeffe_2014}. \citet{bai_2014,bai_2015} also found that the accretion can be largely attributed to angular momentum removal by magnetocentrifugal winds (MCWs). Whereas, when $\bold \Omega \cdot \bold B <0$ accretion is generally suppressed but recently \citet{simon_2015} found bursts of accretion on long timescales attributed to the non-axisymmetric Hall-shear instability. This range of behaviour reflects different physical conditions present at different radii.

Most simulations including the Hall effect are performed using the local shearing box formalism. In this paper, we investigate regions of the disk where different non-ideal effects dominate as a function of radius using global multifluid simulations. This work follows from the work of \citet{okeeffe_2014} by including physically motivated radial density stratification and a radially varying ionisation fraction. We focus on the case where $\bold \Omega \cdot \bold B >0$. 

In this work we only include radial, and not vertical stratification, in order to simplify the process of disentangling the mass accretion due to the MRI (which can be laminar in the case of the Hall effect) and that due to MHD winds. This means our results are of particular relevance for areas close to the midplane of the disk and also more generally, for disks which have experienced dust settling meaning that they are typically flatter.

The paper is laid out as follows: in Section~\ref{sec:numerical} we introduce the multifluid equations used for our simulations and describe the initial setup for all the simulations. In Section~\ref{sec:results} we present our results and we present our conclusions in Section~\ref{sec:conclusions}.

\section{Formulation}
\label{sec:numerical}

\subsection{Multifluid equations}

The numerical simulations presented in this paper were conducted using the multifluid code HYDRA which models weakly ionised plasmas \citep{osullivan_2006,osullivan-2007}. The weakly ionised approximation makes a number of assumptions about the plasma \citep{ciolek_2002,falle_2003}. First, that the mass density is dominated by the neutral species mass density. Second, it assumes that collisions occur predominantly between charged species and neutrals. This allows us to safely neglect the inertia and pressure of the charged species and collisions between charged species. 

These assumptions are reasonable for a PPD where the ionisation fraction is thought to be as low as $\sim10^{-12}$ at 1\,au. The ionisation fraction could be even lower if the dust grains have not settled to the disk midplane. In these simulations we include three fluids: a neutral species, positively charged ions and electrons. The charged particles move in a force free way such that the Lorentz force is balanced by collisions with the neutrals. This allows the ionisation fraction to change spatially and temporally purely due to dynamics without the creation or destruction of the charged species.

The single fluid approximation can be used instead if chemical recombination is faster than all of the relevant dynamic timescales of the system. This implies that a fixed ionisation fraction in time is appropriate. In the case of PPDs, \citet{bai_2011b} justify the single fluid approximation by showing that the orbital frequency is larger than the chemical recombination rate. In Section\,\ref{subsec:dyn-time} we investigate  whether the eddy turn-over time is a more appropriate dynamic timescale to consider for the system and then ultimately whether the single fluid or multifluid approximation is more suitable. 

The multifluid equations implemented in HYDRA are  
\begin{flalign}
\label{eq:multifluid1}
&\frac{\partial \rho_j}{\partial t} + \nabla\cdot{(\rho_j \bm{v}_j)} = 0; \hspace{2mm}(1 \le j \le N) \\
&\frac{\partial \rho_1 \bm{v}_1}{\partial t} + \nabla\cdot{(\rho_1 \bm{v}_1\bm{v}_1 +p_1\bold I)} + \rho_1 \nabla \bm{\Phi} = {\bold J \times \bold B} \label{eq:momentum}\\
&\alpha_j\rho_j(\bold E +\bm{v}_j\times \bold B) + \rho_j\rho_1K_{j,1}(\bm{v}_1 - \bm{v}_j) = 0; \hspace{0.1mm}(2 \le j \le N) \\
&\frac{\partial \bold B}{\partial t} + \nabla\cdot{(\bm{v}_1 \bold B  - \bold B\bm{v}_1)} = -\nabla \times \bold E' 
\label{eq:induction}\\
&\nabla\cdot \bold B = 0 \label{eq:solenoidal}\\
&\nabla \times \bold B = \bold J \\
&\sum\limits_{j=2}^N \alpha_j\rho_j = 0 \\ 
&\sum\limits_{j=2}^N \alpha_j\rho_j \bm{v}_j = \bold J 
\label{eq:multifluid8}
\end{flalign}

\noindent In all of the above equations and throughout the paper the subscript 1, or no subscript, refers to the neutral species. Subsequent subscripts refer to the $N-1$ charged species considered, in our case `2' refers to electrons and `3' to positively charged ions. The number subscripts will sometimes be replaced by letters for clarity, meaning that $1, 2, 3=$n, e, i respectively. The gravitational potential is denoted by $\bm{\Phi}$. The charge-to-mass ratios and collision rates for each of the charged species are denoted by $\alpha_j$ and $K_{j,1}$, respectively.  

The simulations are isothermal, hence the neutral pressure, $p_1$, in Eq.\,\ref{eq:momentum} is calculated from the sound speed, $c_s$, and neutral density, $p_1 = c_s^2 \rho_1$. In Eqs.\,\ref{eq:multifluid1}-\ref{eq:multifluid8} the mass density and velocity of each of the species are denoted by $\rho_j, v_j$ respectively. $\mathrm{\bf{I}}$ is the identity matrix and $c$ is the speed of light. The electric field, magnetic field and current density are denoted by $\mathrm{\bf E, B}$ and $\mathrm{\bf J}$ respectively.

The solenoidal constraint (Eq.\,\ref{eq:solenoidal}) is maintained by hyperbolic divergence cleaning \citep{dedner_2002}. The electric field in the instantaneous rest frame of the neutral fluid, {\bf E$'$} from Eq.\,\ref{eq:induction}, has contributions from each of the non-ideal effects implying
\begin{flalign}
&\bold E^{'}= \left( \eta_\parallel \frac{(\bold J\cdot\bold B) \bold B}{B^2}+ \eta_\mathrm{H}\frac{\bold J \times \bold B}{B} - \eta_\perp \frac {(\bold J
\times \bold B)\times \bold B}{B^2}\right)&
\end{flalign}
\noindent where 
\begin{flalign} 
&\eta_\parallel= \eta_\mathrm{O}=  \frac{1}{ \sigma_\parallel}& \\
&\eta_\mathrm{H}= \frac{\sigma_\mathrm{H}}{ \sigma_{\mathrm{H}}^2 + \sigma_{\perp}^2}& \\
&\eta_\perp     = \frac{\sigma_\perp}{ \sigma_{\mathrm{H}}^2 + \sigma_{\perp}^2}& 
\end{flalign}
\noindent are the parallel (Ohmic), Hall and perpendicular resistivities respectively. Ambipolar resistivity is given by $\eta_\mathrm{A} = \eta_\perp - \eta_\parallel$ and is always positive. The conductivities, $\sigma_\parallel$, $\sigma_\mathrm{H}$ and $\sigma_\perp$, are 
\begin{flalign}
&\sigma_\parallel = \frac{1}{B} \sum \limits_{j=2}^N \alpha_j \rho_j \beta_j & \\ 
&\sigma_\mathrm{H} = \frac{1}{B} \sum \limits_{j=2}^N \frac{\alpha_j \rho_j}{ 1 + \beta_j^2} &\\ 
&\sigma_\perp = \frac{1}{B} \sum \limits_{j=2}^N \frac{\alpha_j \rho_j \beta_j}{1 + \beta_j^2} &
\end{flalign}
where the Hall parameter, $\beta_j$, describes how strongly the charged species are tied to the magnetic field lines. For each of the charged species it is given by,
\begin{flalign}
&\beta_j = \frac{\alpha_jB}{K_{j,1}\rho}
\label{eq:hall-parameter}
\end{flalign}

\subsection{Fluid parameters}
\label{subsec:dim_numbers}
The importance of each non-ideal term relative to the inductive term in the induction equation (Eq.\,\ref{eq:induction}) can be characterised by a number of dimensionless numbers \citep[following][]{balbus_2001,sano_2002a,wardle_2012}. An appropriate characteristic speed ($V$), length ($L$) and diffusion scale ($D$) for the system must be selected. Studies of the linear growth of the MRI \citep[such as][]{wardle_2012} frequently use the Alfv\'en speed, $v_\mathrm{A}$, as the characteristic speed $V$, the orbital period as the characteristic timescale which gives a length scale of \,$L=v_\mathrm{A}/\Omega$ and $D$ is the diffusivity associated with the non-ideal effect in question. 

Therefore, the strength of Ohmic dissipation in comparison to the inductive term in Eq.\,\ref{eq:induction} can be estimated by the following ratio
\begin{equation}
\frac{O}{I} = \frac{D}{VL}  =  \frac{\eta_\mathrm{O} \Omega}{v_\mathrm{A}^2} 
\end{equation}
\noindent where $v_{\mathrm{A}}$ is the Alfv\'en speed given by, 
\begin{equation}
v_{\mathrm{A}} = \frac{B}{\sqrt{4\pi\rho_\mathrm{n}}}
\end{equation}
\noindent Similarly, for the Hall effect and ambipolar diffusion
\begin{equation}
\frac{H}{I} = \frac{\eta_\mathrm{H}\Omega}{v_\mathrm{A}^2} \hspace{3mm} \text{and} \hspace{3mm} \frac{A}{I} = \frac{\eta_\mathrm{A} \Omega}{v_\mathrm{A}^2}
\end{equation}
The relative strength of each of the non-ideal effects can be characterised by examining the following ratios \citep[from][]{wardle_2012}
\begin{equation}
\frac{H}{O} = \frac{\eta_\mathrm{H}}{\eta_\mathrm{O}} \hspace{3mm} \text{and} \hspace{3mm} \frac{A}{H} = \frac{\eta_\mathrm{A}}{\eta_\mathrm{H}}
\end{equation}
\noindent For the Hall effect to be the dominant non-ideal term $\dfrac{H}{O} \gg 1$ and $\dfrac{A}{H}\ll1$. 

\begin{figure}
\centering
 \includegraphics[width=0.5\textwidth]{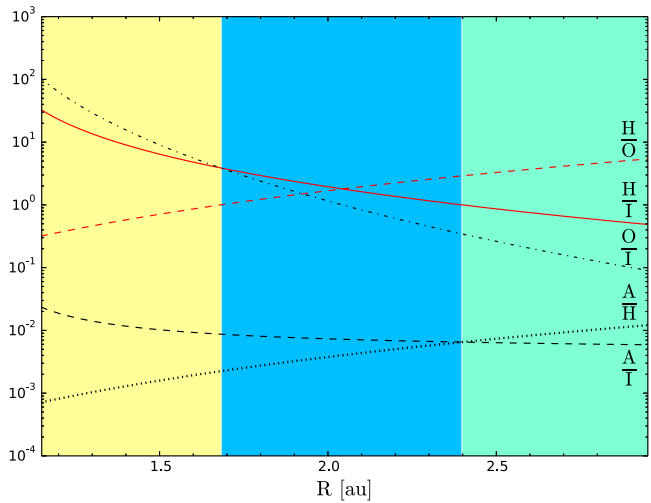}
  \caption{Plot indicating the strength of each of the non-ideal terms relative to each other and to the inductive term as a function of radius: `{\color{red}- -}' is H/O, `$\cdot\, \cdot$' is A/H, `{\bf \color{red}--}' is H/I, `- -' is A/I and `- $\cdot$' is O/I. The yellow region is a strong Ohmic dominated region. The blue region is a strong Hall dominated region and the green region is a weak Hall dominated region. }
\label{fig:elsasser_initial}
\end{figure}

The initial dimensionless numbers for our simulations are shown in Fig.\,\ref{fig:elsasser_initial}. We have divided Fig.\,\ref{fig:elsasser_initial} into three regions: in the yellow innermost region Ohmic dissipation dominates over the Hall effect and the inductive term. We would expect a suppression of the turbulence driven by the MRI in this region if Ohmic dissipation continues to dominate in this region throughout the simulation. The blue region represents where the Hall effect is larger than the inductive term and is also the dominant non-ideal effect. We call this the strong Hall dominated region. Finally, the green region represents where the Hall effect is weaker than the inductive term but remains the dominant non-ideal effect. Even if the Hall effect is weaker than the inductive term it can still affect the nature of the turbulence \citep{downes_2012}. We call this a weak Hall dominated region. The radial extent and strength of these dimensionless numbers evolves during the simulations which is discussed in Section\,\ref{sec:results}.

\subsubsection{Plasma $\beta$}
\noindent An effective plasma $\beta$ (which differs from the Hall parameter, $\beta_j$, defined in Eq.\,\ref{eq:hall-parameter}) can be given by
\begin{flalign}
& \beta  = \frac{\rho c_\mathrm{s}^2}{B_z^2/8\pi}
\end{flalign}
\noindent For these simulations, due to the radial density gradient, $\beta$ varies from $\sim 8.6 \times 10^3 - 2.33 \times 10^5$ at the beginning of the simulations. 

\subsection{Gravity}

A Newtonian gravitational potential is used and given by
\begin{equation}
	\phi = - \frac{GM}{R}
\end{equation}
\noindent where $G$ is the gravitational constant, $M$ the mass of the star taken to be $1\,M_\odot$ and $R=\sqrt{x^2+y^2}$. Note that there is no vertical component of gravity so we are simulating a radially stratified but vertically unstratified disk. Self-gravity in the disk is also neglected.

\subsection{Initial conditions}
\subsubsection{Discussion of Cartesian grid suitability}
We simulate a section of a PPD on a Cartesian grid, the radial extent being either 1\,-\,3.1\,au or 1\,-\,6.1\,au depending on the simulation. Details of the simulations are given in Table~\ref{table:sims}. As discussed in \citet{okeeffe_2014}, our use of a Cartesian grid means that angular momentum is conserved only to the truncation error of our numerical scheme. \citet{okeeffe_2014} investigated this point by plotting the mass accretion rate for the pure hydrodynamic run and for an ideal MHD run. It can be seen that there is negligible accretion in the pure hydrodynamic case (see their Fig.\,4). 

The suitability of different grids was also thoroughly investigated by \citet{valborro_2006} who compared 16 different grid-based (including Cartesian) and smoothed particle hydrodynamics codes. They found that despite the different methods all the codes gave generally the same results. More recently \citet{lyra_2008} performed ideal MHD simulations on an Cartesian grid and also concluded that Cartesian grids are suitable for accretion disk problems. 

\subsubsection{Boundary conditions}
\label{subsubsection:bc}
There are three types of boundary conditions needed for a $\pi/2$-global accretion disk model which uses Cartesian coordinates. The same boundary conditions are used as in \citet{okeeffe_2014}. 

Briefly to recapitulate, since we are only simulating a quarter of the disk, for the vertical box-faces material flowing off the grid must be fed back into the disk to mimic a full $2\pi$ disk. The second type of boundary condition is for the $z$-boundaries which are periodic.

Last are the interior boundaries which deal with the cylindrical shape of the accretion disk enclosed in the square prism which represents the computational grid. When $R < R_\mathrm{in} $ and $R > R_\mathrm{out}$ the physical variables are not updated during the simulations, these regions are called frozen zones. An inner and outer buffer zone, called wavekilling regions, are also implemented. These are used to smooth the transition between the active computational domain and the frozen zones, helping to damp unphysical waves that would otherwise reflect from internal boundaries. In these wavekilling regions variables are driven back to their initial values at a rate which is dependent on a driving function, S(R), (see \citet{okeeffe_2014} and \citet{lyra_2008} for more details). These wavekilling regions are particularly well suited to damp waves whose wavelengths are either equal to the width of the wavekilling region or less. For the 3\,au runs, the inner wavekilling region is located between $1.0-1.1\,$au and the outer wavekilling region is between $3.0-3.1\,$au. For the 6\,au runs the outer wavekilling region is between $6.0-6.1\,$au.

We also apply an inward radial flow in the wavekilling regions, following \citet{fromang_2006}, so that material does not accumulate (dissipate) at the inner (outer) boundary. We similarly take $v_\mathrm{R} = \dfrac{-3\alpha c_s^2}{2\Omega}$, with $\Omega = \sqrt{\dfrac{GM}{R^3}}$ and $\alpha=7.5\times10^{-2}$ . This value for $\alpha$ was chosen as it was the value of $\alpha$ found by \citet{okeeffe_2014} whose simulations are the most similar to those presented here. It is also within the acceptable range of values obtained when comparing with observations.

\subsubsection{Physical parameters}
The disk is set up initially in hydrostatic equilibrium. The temperature is taken to be 280\,K for all the simulations which is taken from Table\,1 from \citet{salmeron_2003}. The initial magnetic field is set to be 100\,mG and is introduced at the beginning of the simulations. 

As in \citet{okeeffe_2014}, the average mass of the neutral particles is taken to be $m =  2.33m_\mathrm{p}$ (where $m_\mathrm{p}$ is the mass of the proton). The ion fluid is meant to represent the average of a number of metal atoms which can be modelled as a single atom \citep{umebayashi_1990}. The average mass of the ions is therefore taken to be $m_{\mathrm{ion}} = 24m_\mathrm{p}$. The negatively charged fluid is taken to be electrons. 

The charge-to-mass ratios for the electrons and the ion species are $\alpha_\mathrm{e} = -5.28\times 10^{17}\mathrm{statC\,g^{-1}}$ and $\alpha_\mathrm{i}=1.20\times 10^{13}\mathrm{statC\,g^{-1}}$. Similar to \citet{okeeffe_2014}, we use the rate coefficients for momentum transfer by elastic scattering of the charged species with neutrals given in \citet{wardle_ng_1999}. The collision rates, $K_{j,1}$, for each of the charged species are $K_\mathrm{e,n}=2.43\times 10^{15} \mathrm{cm^{3}g^{-1}s^{-1}}$ and $K_{\mathrm{i,n}} = 3.64\times 10^{13} \mathrm{cm^{3}g^{-1}s^{-1}}$. 

\noindent We have introduced an initial radial density profile for the neutral species. The neutral mass density was calculated to approximate the values given in Table\,1 from \citet{salmeron_2003} for 1, 5 and 10\,au and is given simply by $\rho(R) = \rho_0\left(\dfrac{R_\mathrm{in}}{R}\right)^3$ with $\rho_0=2.33\times 10^{-9} \mathrm{g\,} \mathrm{cm}^{-3}$ and $R_\mathrm{in}=1$\,au. To set the simulations up in hydrostatic equilibrium the initial azimuthal velocity field differs from Keplerian rotation (due to the pressure gradient introduced by varying the density radially) in the following way,
\begin{equation}
v_\phi(R)=\sqrt{\frac{GM}{R} - c_s^2\frac{R_\mathrm{in}}{R}}
\end{equation}

We have also introduced a more physical ionisation equilibrium which varies as a function of radius. We fit values taken from Table\,1 in \citet{salmeron_2003} with a quadratic function. These values include ionisation processes by cosmic rays, radioactive elements \citep{umebayashi_1981} and X-rays from the protostar \citep{igea_1999} balanced by recombination processes on the surface of dust grains and also in the gas phase \citep{nishi_1991}. The dust grains are assumed to have settled out \citep{fromang_2002}. This means $\rho_\mathrm{e}(R) = f(R)\rho$ where $f(R) = 3.1716\times10^{-15}R^2+ 1.0782\times10^{-14}R -1.3783\times 10^{-14}$. Due to quasi-neutrality, $\rho_{\mathrm{i}}(R) = -\dfrac{\alpha_\mathrm{e}\rho_\mathrm{e}(R)}{\alpha_{\mathrm{i}}}$. The ionisation fraction is not fixed in time or space in the simulations, meaning it can change as long as local charge neutrality is maintained. 

\subsection{Grid setup considerations}
\citet{balbus_1991} used linear analysis to show that, in order to study the MRI, the critical vertical wavelength that must be resolved is given by
\begin{equation}
\lambda_c =\frac{2\pi}{\sqrt{3}}\frac{v_\mathrm{Az}}{\Omega}
\label{eq:lambda_crit}
\end{equation}
\noindent \noindent where $v_{\mathrm{Az}}$ is the Alfv\'en speed for a vertical magnetic field. In the simulations presented here the critical wavelength is initially resolved by $\sim$22 grid zones for the 3\,au runs and by $\sim$5 grid zones for the 6\,au runs. 

Including vertical stratification may result in the critical wavelength being larger than the vertical extent of the box at small radii which would be likely to result in different behaviour in the inner regions of the disk than presented here. The vertical extent of the disk increases sharply with increasing radius so this possibility is of particular concern in the very inner regions of the disk. It is still possible that turbulence could arise in these regions due to interactions with turbulent neighbouring material but may take longer to develop. The inner regions of PPDs remain largely unresolved by observations making it difficult to determine whether indeed the vertical extent of the disk is as small as obtained by calculating the pressure scale height by assuming hydrostatic equilibrium.

\subsection{Diagnostics}
\subsubsection{Angular momentum transport}

One of the most common quantities used to determine the amount of angular momentum transport occuring in PPDs is the dimensionless viscous stress parameter, $\alpha$, introduced by \citet{sunyaev_1973}. The viscous stress parameter has contributions from the Reynolds stress and the Maxwell stress weighted by the pressure which are denoted as $\alpha_\mathrm{R}$ and $\alpha_\mathrm{M}$ respectively. 

\noindent The Reynolds and Maxwell stresses are given by
\begin{equation}
T_\mathrm{R} =\rho\delta v_\mathrm{R} \delta v_\phi \hspace{1mm}; \hspace{3mm} T_\mathrm{M} =  -B_\mathrm{R} B_\phi/4\pi
\end{equation}

\noindent The fluctuating velocity components are  $\delta v_i=v_i - v_{i,\mathrm{kep}}$, where $i=R,\phi$. In the following, angle brackets are used to denote a volume-averaged quantity and time-averaging of quantities is denoted by $\overline{\, \cdot \,}$. Following \citet{flock_2011}, we calculate $\langle \alpha \rangle$ by integrating the mass-weighted stresses over the total domain,
\begin{align}
\langle \alpha\rangle &= \langle \alpha_\mathrm{R}\rangle +\langle \alpha_\mathrm{M} \rangle \nonumber \\
 &=\frac{1}{c_s^2}\left(\frac{\int T_\mathrm{R} dV}{\int \rho dV} -\frac{\int T_\mathrm{M} dV} { \int \rho dV}\right)
\end{align}
\noindent and $\overline{\alpha}$ is calculated by,
\begin{equation} 
\overline{\alpha} = \overline{\alpha_\mathrm{R}} + \overline{\alpha_\mathrm{M}}
\end{equation}
\subsubsection{Mass accretion rate}
Typically, for numerical simulations the viscous stress parameter is used to quantify the efficacy of angular momentum transport, it being a local property of the disk. In contrast, observations of UV excess emission \citep[for instance,][]{manara-2014} instead provides a way to infer the mass accretion rate, $\dot{M}$, onto the central protostar. This measure cannot directly determine accretion in the disk itself. To compare with observations we estimate $\dot{M}$ as well as $\alpha$. We calculate $\dot{M}$ per pressure scale height, $h$, in the vertical direction. The pressure scale height is given by
\begin{equation}
h(R) = \frac{c_s}{\Omega}
\end{equation}
\noindent and for the given parameters we have chosen, $h(1.1\,\mathrm{au})\sim0.035$\,au. The mass accretion rate, at a particular radius $R$ and between a height $s$ above and below the disk midplane, can be calculated as follows,
\begin{equation}
\dot{M} = - \int \limits_{0}^{2\pi}\int \limits_{-s}^{+s}\rho v_\mathrm{R} Rdz d\phi 
\end{equation}
To calculate an accretion rate per scale height we integrate over the entire height of the simulation box and divide by the number of scale heights in the vertical direction.

\subsection{Description of simulations}
The results of this paper are based on 4 simulations: 2 multifluid and 2 ideal MHD simulations. The details of the simulations are given in Table\,\ref{table:sims}. We aim to compare the multifluid and ideal simulations to ascertain the specific differences that the inclusion of all 3 non-ideal effects generate. The radial extent for the multifluid simulations changes from 3\,au (mf-3au) to 6\,au (mf-6au), the same applies for the ideal MHD runs (ideal-3au and ideal-6au). The length of the simulations are given in units of orbital periods defined at 1\,au and differ due to computational resources. In general, the simulations were run for twice as long as those in \citet{okeeffe_2014} and \citet{okeeffe_2015} to more thoroughly study the non-linear phase of the MRI. The derived quantities given in Table\,\ref{table:sims} are described in the following sections. In order to avoid the effect of initial transients time averaging is not performed from the beginning of the simulations. Instead, all time averaging is performed between 45-66\,(35-123) orbits for the 3\,(6)\,au runs. Similarly, all volume averaging in the radial direction is performed slightly interior to the wavekilling regions between 1.15-2.95\,(1.15-5.95)\,au for the 3\,(6)\,au runs. 
\begin{figure*}[H]%
	\centering
    \subfigure[ ]{%
        \includegraphics[width=0.5\textwidth]{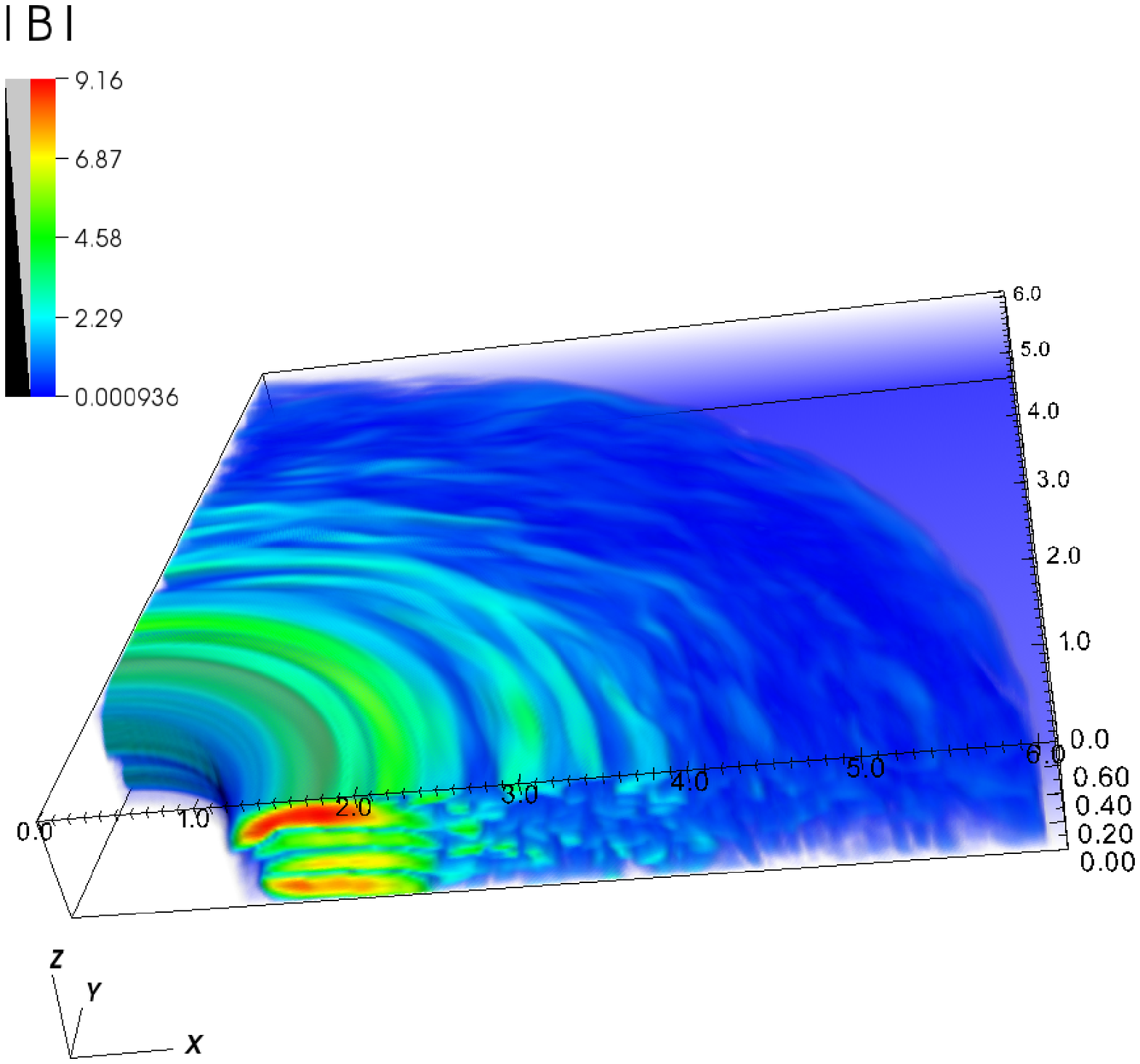}
       	\centering
\label{fig:visit-b-mf}}%
  	    ~
    \subfigure[ ]{%
        \includegraphics[width=0.5\textwidth]{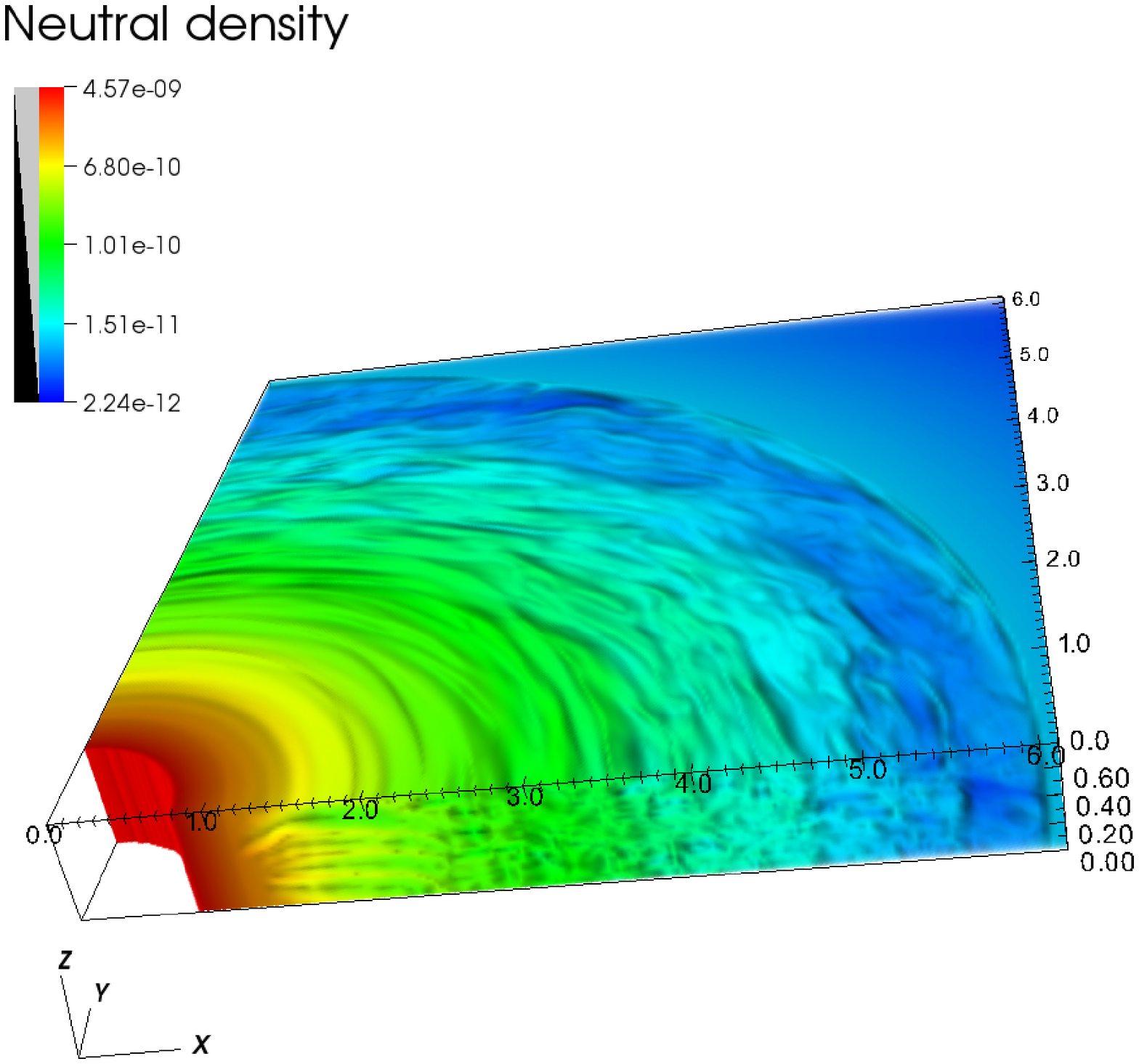}
	\centering
 \label{fig:visit-rho-mf}}%
    \caption{(a) plots the magnitude of the magnetic field and (b) plots the log of the neutral density for mf-6au. The units for $|\bf{B}|$ are Gauss and the units for $\rho$ are $\mathrm{g\, cm^{-3}}$.} 
    \label{fig:visit-mf}%
\quad    
    \subfigure[ ]{%
        \includegraphics[width=0.5\textwidth]{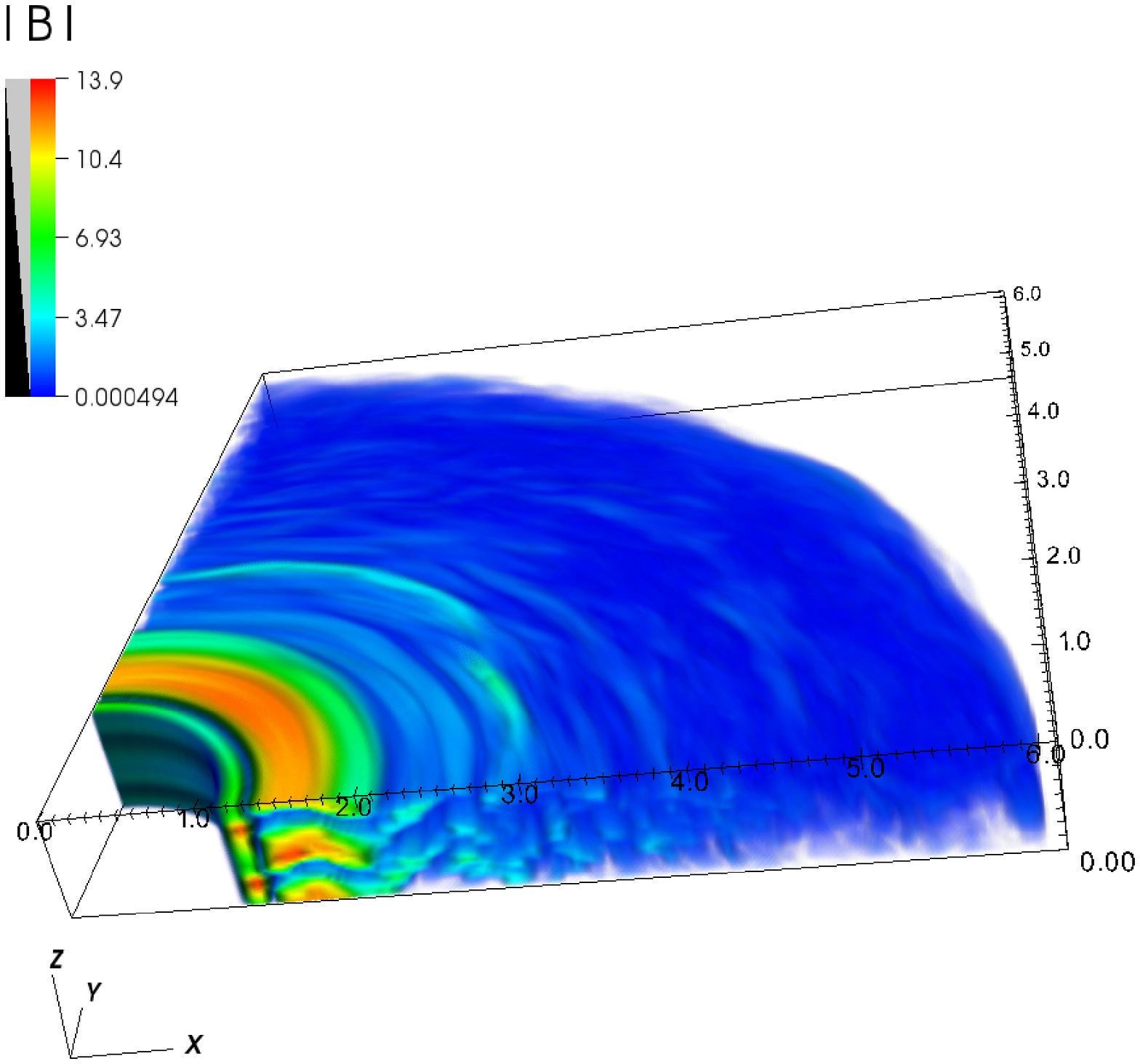}
  	    \label{fig:visit-b-ideal}}%
  	    ~
    \subfigure[ ]{%
        \includegraphics[width=0.5\textwidth]{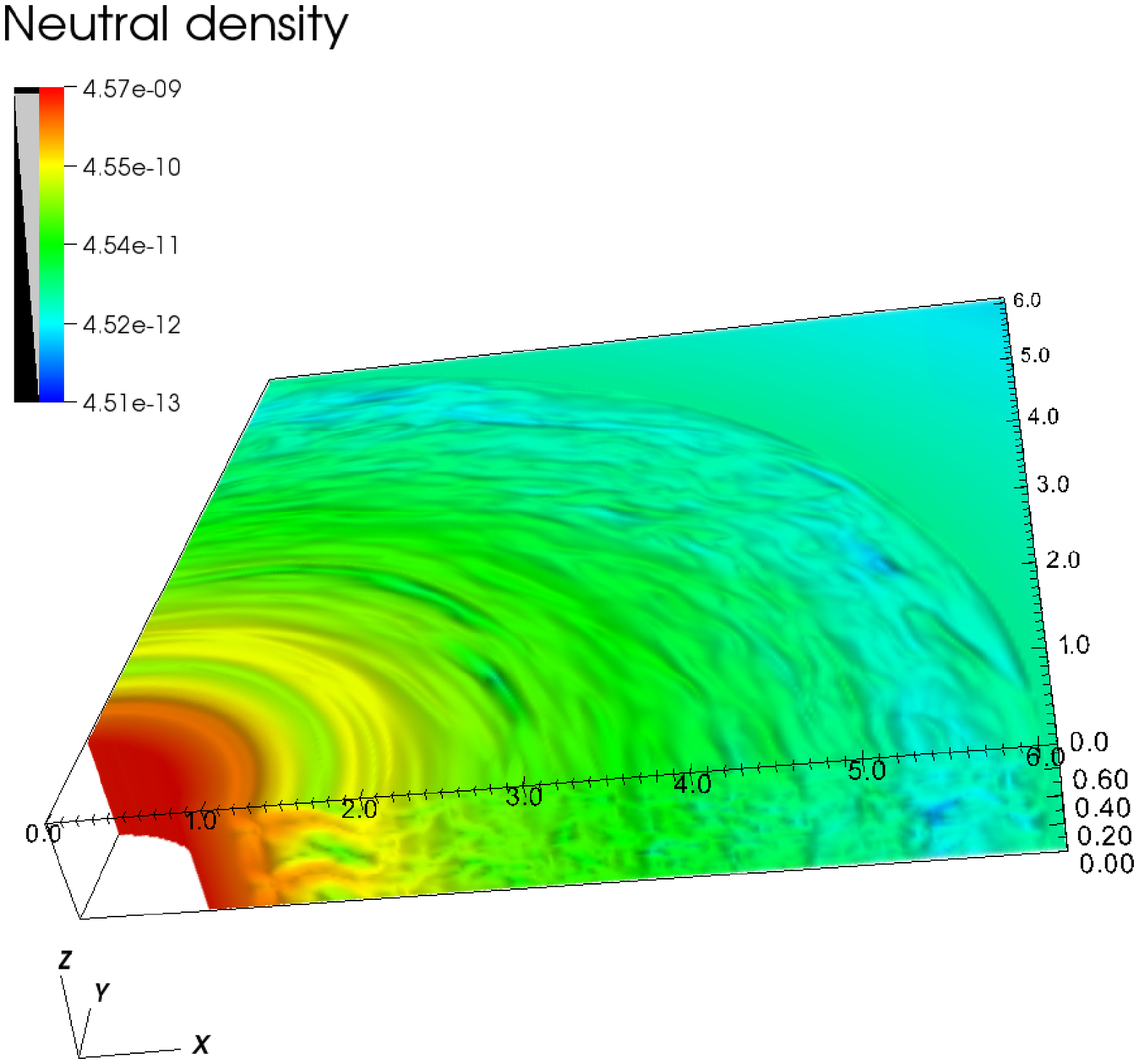}
		\label{fig:visit-rho-ideal}}
    \caption{(a) plots the magnitude of the magnetic field and (b) plots the log of the neutral density for ideal-6au. Units are as in Fig.\,\ref{fig:visit-mf}. Note that the colour scales are not the same as for Fig.\,\ref{fig:visit-mf}.} 
    \label{fig:visit-ideal} 
\end{figure*}

\begin{table*}
\centering
\caption{Summary of simulation parameters and results. The uncertainties quoted for $\langle \overline{\alpha} \rangle$ represent the standard deviation of the data between the number of orbits given in column 6. In column 6 and 7, T.A. and V.A. stand for time averaging and volume averaging, respectively. }
	\begin{tabular}{@{}lccccccccclll@{}}
		\hline
		Run ID & $N_{x,y}$ & $N_z$ & $x,y$ & $z$ & T.A. & V.A. & $\langle \overline{\alpha} \rangle$& $\overline{\dot{M}}$ & $\langle \overline{|B|} \rangle$  & $\langle \overline{\alpha_\mathrm{RM}} \rangle$ & $\langle \overline{\alpha_\mathrm{RM}} \rangle$ \\
		 &  & & [au]& [au] & [orbits] & [au] & &[$M_\odot\,\mathrm{yr^{-1}}\,h^{-1}$] & G & inner & outer\\
		\hline
	mf-6au 	 & 512	& 64   & 6.1 & 0.8 & 35-123 & 1.15-5.95 & 0.017 $\pm$ 0.005 & 2.2 $\times 10^{-7}$ & 1.6 & 0.09 & 0.64 \\
	mf-3au    & 512  & 128  & 3.1 & 0.8 & 45-66  & 1.15-2.95 & 0.010 $\pm$ 0.001 & 1.3 $\times 10^{-6}$ & 2.8 & 0.13 &0.39 \\
	ideal-6au & 512	& 64   & 6.1 & 0.8 & 35-123 & 1.15-5.95 & 0.012 $\pm$ 0.002 & 2.0 $\times 10^{-7}$ & 0.8 & 0.21 & 0.63\\
	ideal-3au & 512	& 128  & 3.1 & 0.8   & 45-66  & 1.15-2.95 & 0.006 $\pm$ 0.001 & 6.5 $\times 10^{-7}$ & 2.1 & 0.25 &0.47\\
		\hline 
	\label{table:sims}
	\end{tabular}
\end{table*}

\section{Results}
\label{sec:results}

\subsection{Structure of the magnetic field}
\label{subsec:mag-struc}

In both mf-6au and ideal-6au stable magnetic structures form in the inner regions of the disk. These features can be seen in Fig.\,\ref{fig:visit-b-mf} and Fig.\,\ref{fig:visit-b-ideal} which plot $|\bold B|$ at $t=123$\,orbits for mf-6au and ideal-6au, respectively. This structure occurs in regions of the disk where $|\bold B|$ is largest for both simulations. The magnetic field becomes strongest in the inner regions because the angular velocity is largest in this region of the disk, creating toroidal field from the winding of the vertical magnetic field. 

The structure is noticeably different between mf-6au and ideal-6au which can be seen by comparing Fig.\,\ref{fig:visit-b-mf} and Fig.\,\ref{fig:visit-b-ideal}. This would suggest that the dominant non-ideal effect in this region, namely the Hall effect, is connected with this difference in behaviour. For ideal-6au there are fewer, but stronger, alterations between strong and weak $|\bold B|$ than for mf-6au. Different behaviour is seen for mf-6au because the Hall effect twists, and therefore disorders, the magnetic field. 

\begin{figure}
\centering
 \includegraphics[width=0.5\textwidth]{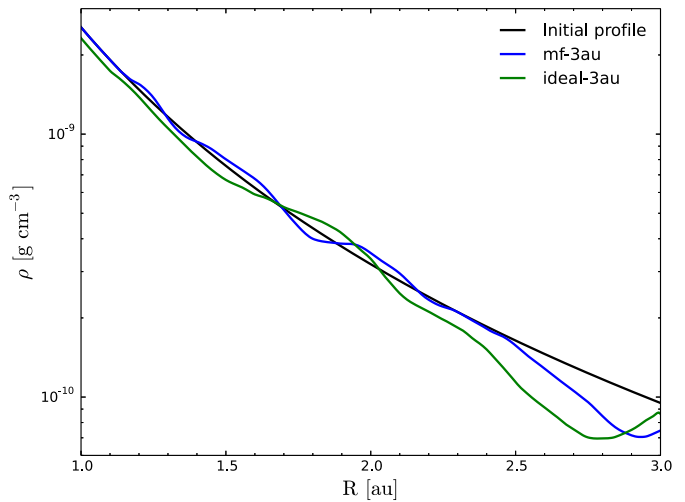}
  \caption{Initial neutral density profile (black dashed line), overplotted is the neutral density, vertically and time-averaged, for mf-3au and ideal-3au.}
\label{fig:neutral-density}
\end{figure}

Similar structure is also visible in the density plots for mf-6au and ideal-6au (see Figs.\,\ref{fig:visit-rho-mf} and \ref{fig:visit-rho-ideal}) except that areas of strong magnetic field correspond to areas of low density and vice versa. In contrast, the time and vertically averaged densities, shown in Fig.\,\ref{fig:neutral-density}, do not display either a marked under- or over-density coincident with the magnetic field structure.

Similar, yet not so pronounced, magnetic and density structures were observed in the higher resolution runs, mf-3au and ideal-3au. The 6\,au simulations were run for 123 orbits whereas the 3\,au runs were only run for 66 orbits. This may explain why the structure is not as pronounced for the 3\,au runs as it has not had time to fully develop. In fact, at 66 orbits for the 6\,au runs the structure is similar to what was seen at the same time for the 3\,au runs.

\subsection{Angular momentum transport and dimensionless numbers}
\label{subsec:alpha}
\begin{figure}
\centering
 \includegraphics[width=0.5\textwidth]{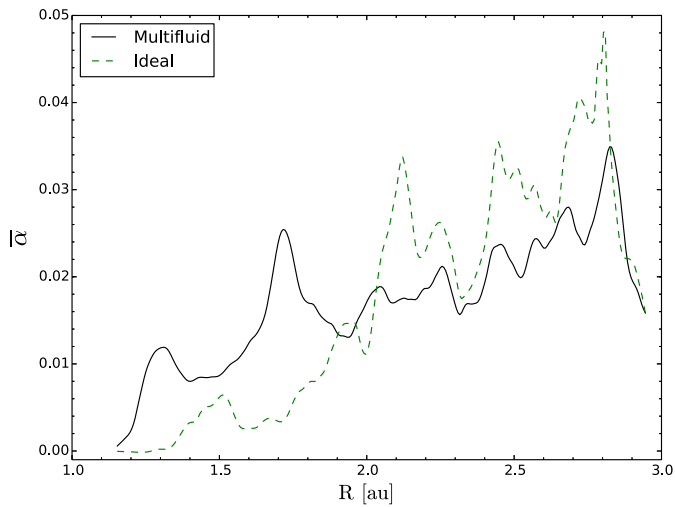}
  \caption{Comparison of time-averaged $\alpha$ for ideal-3au and mf-3au. }
\label{fig:alpha_r_3au_mf_ideal}
\end{figure}

Next, we investigate how this magnetic structure affects angular momentum transport. Fig.\,\ref{fig:alpha_r_3au_mf_ideal} contains a plot of $\overline{\alpha}$ as a function of radius, both time and vertically averaged, for mf-3au and ideal-3au. In the inner region of the disk (out to $\sim$2\,au) $\overline{\alpha}$ is larger for mf-3au than ideal-3au. Similar behaviour was again seen in the 6\,au simulations except $\overline{\alpha}$ is larger for mf-6au out to $\sim$3\,au instead.

\begin{figure}
\centering
 \includegraphics[width=0.5\textwidth]{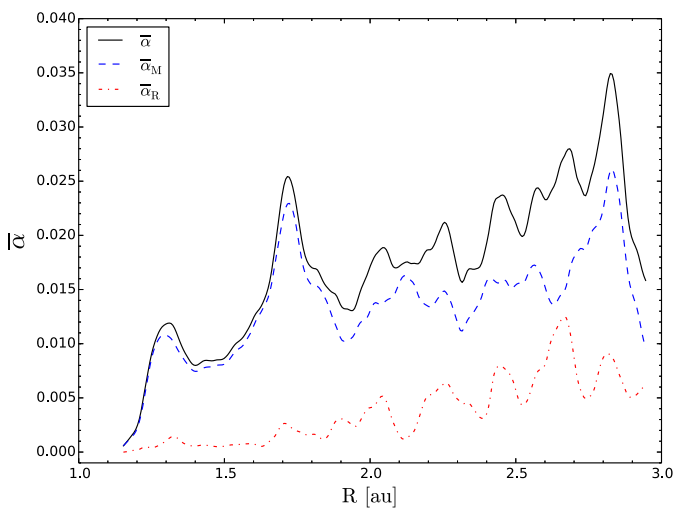}
  \caption{Time-averaged plot of $\alpha, \alpha_\mathrm{R}, \alpha_\mathrm{M}$ plotted as a function of radius for mf-3au.}
\label{fig:alpha_r_3au_mf}
\end{figure}

By examining the separate components of $\overline{\alpha}$ for mf-3au (shown in Fig.\,\ref{fig:alpha_r_3au_mf}) we can identify the cause of this behaviour. The contribution from the Maxwell stresses, given by $\overline{\alpha}_\mathrm{M}$, to the overall value of $\overline{\alpha}$ in the inner regions of the disk is significant, with very little attributable to the Reynolds stress component, given by $\overline{\alpha}_\mathrm{R}$. This indicates that while accretion is occuring the flow is mainly laminar in the inner parts of the disk which was seen in \citet{lesur_2014}. In contrast, for ideal-3au $\overline{\alpha}_\mathrm{M}$ does not increase in the inner parts of the disk.

\begin{figure}
\centering
 \includegraphics[width=0.5\textwidth]{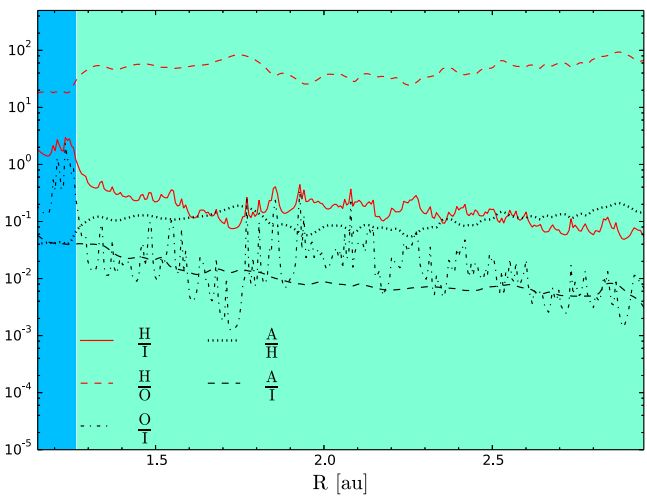}
  \caption{The dimensionless numbers described in Section~\ref{sec:numerical} are shown as a function of radius time-averaged between 45-66 orbits for mf-3au. The linestyles represent the same quantities shown in Fig.\,\ref{fig:elsasser_initial} and described in Section\,\ref{subsec:dim_numbers}. Similar behaviour is seen for mf-6au.}
\label{fig:elsasser066}
\end{figure}

Now by comparing Fig.\,\ref{fig:elsasser066}, which plots the time-averaged dimensionless numbers for mf-3au, with the initial dimensionless numbers (Fig.\,\ref{fig:elsasser_initial}) the most obvious difference is the strength and radial influence of the Hall effect. For both of the multifluid simulations the Hall effect has become weaker at smaller radii, only dominating the inductive term out to a radius of $\sim$1.25\,au instead of between $\sim 1.7-2.4$\,au for mf-3au. For mf-6au the outer radius of the strong Hall dominated region also decreases from $\sim$2.4\,au to $\sim$1.7\,au. 

The dimensionless number $\dfrac{H}{I}$ (representing the strength of the Hall effect in comparison to the inductive term described in Section\,\ref{subsec:dim_numbers}) decreases by roughly an order of magnitude for mf-3au. For mf-6au, the dimensionless numbers evolve in a similar fashion except that $\dfrac{H}{I}$ is larger at small radii than for mf-3au, increasing to $\sim10$ at 1.2\,au.

Examining changes in the strength of the other non-ideal effects shows that $\dfrac{A}{H}$ and $\dfrac{H}{O}$ (representing the strength of ambipolar diffusion and Ohmic dissipation in comparison to the Hall effect respectively) both increase throughout the disk. The dimensionless number $\dfrac{A}{H}$ still remains less than unity.

\subsection{Density profile and ionisation fraction}
All four simulations, irrespective of the difference in radial extent, show an under-density at the outer boundary of the disk (see Fig.\,\ref{fig:neutral-density}). This would indicate that material is being accreted from the outer parts of the disk faster than it is replenished by the inward flow that we implement in the wavekilling regions.

\begin{figure}
\centering
 \includegraphics[width=0.5\textwidth]{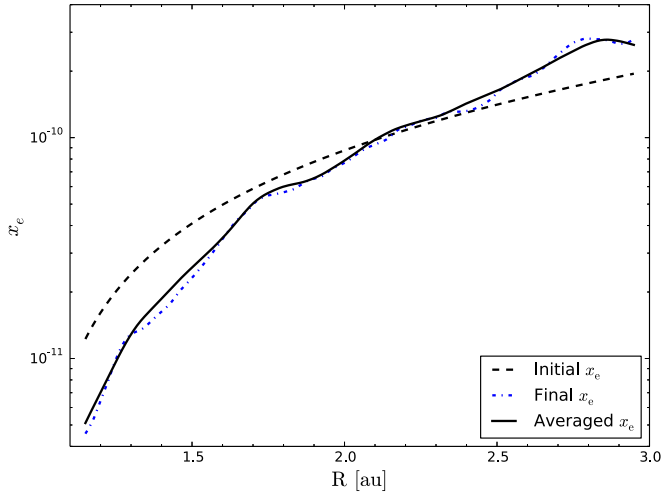}
  \caption{Initial ionisation fraction profile (black dashed line) as a function of radius, overplotted is the ionisation fraction after 66 orbits (blue dash-dotted line) for mf-3au and the average ionisation fraction (black solid line) between 45-66 orbits.}
\label{fig:ionisation-fraction}
\end{figure}

The ionisation fraction profile in the disk changes by less than a factor of two in comparison to the initial profile in most parts of the disk, shown in Fig.\,\ref{fig:ionisation-fraction}. Single fluid simulations assume the ionisation fraction of the disk can be calculated from the neutral density and that kinematics in the system do not change the ionisation fraction. Our results showing that the ionisation fraction does not change significantly suggests that these multifluid simulations should be comparable to single fluid simulations. By comparing Fig.\,\ref{fig:ionisation-fraction} with the neutral density plot (Fig.\,\ref{fig:neutral-density}), the ionisation fraction can be seen to have decreased in areas where the neutral density is higher and vice versa. The ion and electron densities have also changed though and are similar to the neutral density profile, they merely have not changed quite as much. This implies that the charged species are not accreting as easily as the neutral fluid and yet they mediate the accretion process. Overall, the variations from the initial ionisation fraction are much less than an order of magnitude at all radii.

\subsection{Turbulent and laminar flow}
\label{subsec:turb-lam}

Global simulations are required in order to study the interaction between regions of PPDs governed by different physical processes. Planet formation is unlikely to occur in areas with high levels of turbulence making the characterisation of the physical properties of PPDs at different radii an important task. As mentioned in Section\,\ref{subsec:alpha}, the Maxwell stresses dominate the contribution to $\alpha$ in the inner regions of the disk, out to $\sim$2\,au, for mf-3au in comparison to ideal-3au. This region appears to separate laminar and turbulent flows, as seen in Fig.\,\ref{fig:visit-rho-mf}.
 
It can be reasonable to replace the detailed description of non-ideal MHD used here with hydrodynamics by including a viscous term scaled by the value of $\alpha$ (this is much less computationally demanding than MHD simulations) if the outcome of global non-ideal MHD simulations resulted in turbulent transport. On the other hand, if accretion occurs via laminar rather than turbulent flow this approximation no longer holds. The simulations presented here \citep[and others such as those of][]{lesur_2014} display a mixture of laminar and turbulent flows (see Fig.\,\ref{fig:visit-rho-mf}) making it important to characterise the flow by examining the ratio of the two contributions to $\alpha$, namely $\alpha_\mathrm{R}$ and $\alpha_\mathrm{M}$. This ratio has been investigated before by \citet{fromang_2006} for ideal MHD simulations and was found to be approximately 1:3. We introduce a specific dimensionless number, $\alpha_\mathrm{RM}$, to describe this ratio,
\begin{equation}
\alpha_\mathrm{RM} = \frac{\alpha_\mathrm{R}}{\alpha_\mathrm{M}}
\end{equation}

For our simulations, and only considering areas of significant accretion ($\alpha \ge 7 \times 10^{-4}$), we examine this ratio. We define the boundary between structured magnetic field and more turbulent flow by examining the point at which the Maxwell stresses dominate the contribution to $\alpha$ for the multifluid simulations in comparison to the ideal MHD simulations. For the multifluid 3\,(6)\,au simulation this boundary occurs at 2.1\,(2.8)\,au. We average $\overline{\alpha_\mathrm{RM}}$ between 1.15-2.1\,(1.15-2.8)\,au and 2.1-2.95\,(2.8-5.95)\,au for the 3\,(6)au runs, centring these regions around the transition region.

For mf-3au in the inner region of the disk we found $\overline{\alpha_\mathrm{RM}}=0.13$ and for the outer turbulent region $\overline{\alpha_\mathrm{RM}}=0.39$. In contrast, for ideal-3au $\overline{\alpha_\mathrm{RM}}=0.25$ for the inner region and $\overline{\alpha_\mathrm{RM}}=0.47$ for the outer region. Similar behaviour is found for the 6\,au simulations, the values of $\overline{\alpha_\mathrm{RM}}$ obtained are given in Table\,\ref{table:sims}. We find that $\alpha_\mathrm{RM} \lesssim 0.1$ corresponds to accretion via strong Maxwell stresses. This dimensionless number can be used together with $\alpha$ to ascertain the nature of the accretion. 

By examining accretion across this boundary it can be seen from Fig.\,\ref{fig:alpha_r_3au_mf_ideal} that there is a continual decrease in angular momentum transport with decreasing radius for ideal-3au. However, while similar behaviour is observed in mf-3au, in this system there is still significant angular momentum transport at small radii. This is due to the increase in the Maxwell stresses for mf-3au (shown in Fig.\,\ref{fig:alpha_r_3au_mf}) which is not present for ideal-3au. The overall density fluctuations for any of the simulations from the initial values are less than a factor of two at the boundary between more structured magnetic and density features and turbulent flow (see Fig.\,\ref{fig:neutral-density}) so no gaps exist. Despite this, small scale feautures do exist (see Fig.\,\ref{fig:visit-b-mf} and Fig.\,\ref{fig:visit-b-ideal}).

\subsection{Dynamic timescale}
\label{subsec:dyn-time}
We are interested in estimating whether the eddies created by MRI driven turbulence can change the ionisation fraction by mixing ionised material either vertically or radially in the disk. This would potentially allow magnetically dead regions of the disk to be revived. For this to be possible the eddy turn-over time must be shorter than the chemical recombination timescale. 

Separately, this has implications for the physical assumptions that should be implemented in the numerical method used to model a PPD. The single fluid approximation is appropriate when chemical recombination is faster than any of the dynamic timescales associated with the system. \citet{bai_2011b} showed that the recombination time is much shorter than the orbital timescale. 

\subsubsection{Comparison of Keplerian orbital timescale with eddy turn-over time}
In incompressible fluid turbulence the ``eddy turn-over time'' can be thought of as the time it takes for a fluid element to circulate fully around a vortex of a particular size.  Although in compressible MHD turbulence this interpretation is no longer correct, the eddy turn-over time is still a useful indicator of the turbulent timescale at a particular lengthscale. Thus, the presence of turbulence introduces a range of new timescales into the dynamics of accretion disks.

To determine the appropriate timescale to use as the dynamic timescale for PPDs  we compare the orbital timescale at 1\,au with the eddy turn-over times, $t_\mathrm{eddy}(k)$, (where $k$ is the wavenumber), as they are a measure of the timescales for turbulent mixing. To calculate $t_\mathrm{eddy}(k)$, first, we subtract the initial velocity profile from the velocities, leaving the turbulent velocities
\begin{equation}
\delta v_i = v_i - v_{i,\mathrm{kep}}
\end{equation}

\noindent where $i=R,\phi$. We take the power spectra of the different components of $\delta v_i$ allowing us to examine the characteristic velocities, $v_\mathrm{eddy}(k)$, as a function of lengthscale. These velocities are averaged over time. We study the power spectra of the radial and vertical components of $\delta v_i$ separately to examine any directional dependence due to the anisotropic nature of the turbulence produced by the MRI \citep{murphy-2015}.

\noindent In terms of the power spectrum, $P(k)$, $v_\mathrm{eddy}$ is given by
\begin{equation}
v_\mathrm{eddy}(k) = \sqrt{2P(k)}
\end{equation}

\noindent and we can then express $t_\mathrm{eddy}(k)$ as
\begin{equation}
t_\mathrm{eddy}(k)= \frac{2\pi}{kv_\mathrm{eddy}(k)}
\end{equation}

In mf-3au the Maxwell stresses vary strongly as a function of radius, especially in comparison to ideal-3au, which led us to investigate if there were noticeable differences in $t_\mathrm{eddy}$ as a function of radius. We calculated $t_\mathrm{eddy}$ as described above for two different boxes: one located between 1.31-2.1\,au and the other between 2.1-2.87\,au. \citet{downes_2012} determined for HYDRA that numerical dissipation occurs on length scales less than 10-15 grid zones so we cannot examine the power spectra on length scales less than this.

For mf-3au, $\Omega t_\mathrm{eddy} \gg 1$ for all wavenumbers, as shown in Fig.\,\ref{fig:teddy-mf} indicating that the orbital timescale is much shorter than the eddy turnover time. This implies that the orbital timescale is the more appropriate timescale to consider when comparing with the recombination timescale, in order to determine whether single fluid or multifluid simulations are more appropriate. The results are similar for mf-6au and also for both of the ideal MHD simulations.

\begin{figure}
\centering
 \includegraphics[width=0.5\textwidth]{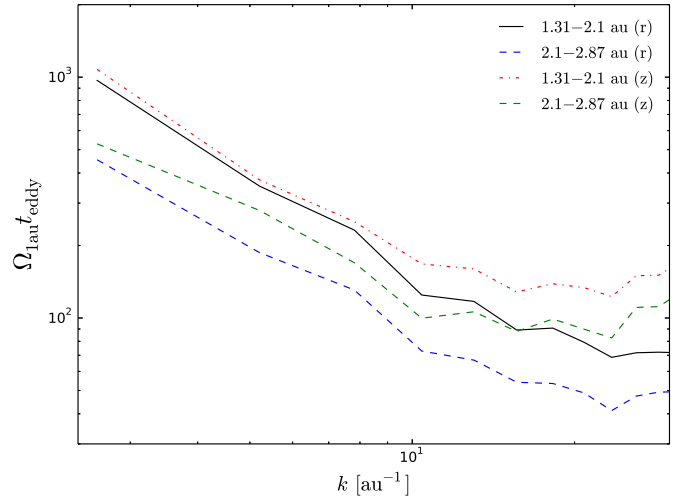}
  \caption{The Keplerian frequency at 1\,au multiplied by the eddy turnover time is plotted as a function of wavenumber, $k$, for mf-3au. The solid black and blue dashed lines show $\Omega_\mathrm{1au} t_\mathrm{eddy}$ for an inner and an outer radial region using radial turbulent velocities. The red dash-dotted and green dashed lines show the same quantity but the vertical turbulent velocities are used instead. }
\label{fig:teddy-omega-mf}
\end{figure}

There is also little difference for mf-3au between the power spectra taken at different radii, only a vertical shift which is most probably linked to the different Keplerian velocities at the different radii. We also examine the difference found by comparing power spectra for the radial and vertical directions. For all the simulations there are slightly larger eddy velocities, for all wavenumbers, in the radial rather than the vertical direction. This is to be expected since the MRI creates turbulence due to the Keplerian shear in the radial direction. Therefore, turbulence is preferentially driven in the radial direction.   

\subsubsection{Comparison of an effective recombination timescale with eddy turn-over time}

We then use the effective recombination time, $t_\mathrm{rcb}$, calculated in \citet{bai_2011b} to compare with $t_\mathrm{eddy}$. This recombination time is sensitive to the fastest recombination process occuring at $\sim$1\,au near the midplane of the disk. From Fig.\,2 of \citet{bai_2011b} we consider the top left panel representing parameters for a PPD at 1\,au not including grains which is most similar to our simulations. At $z/h < 1$, $\mathrm{log}_{10}(\Omega t_\mathrm{rcb})=-1$ corresponding to $t_\mathrm{rcb}=5.0\times 10^{5}\,\mathrm{s}$ at 1\,au.

\begin{figure}
\centering
 \includegraphics[width=0.5\textwidth]{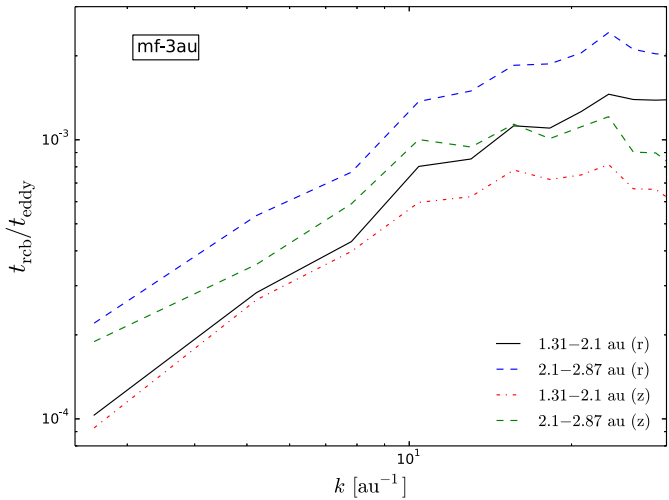}
  \caption{$t_\mathrm{eddy}/t_\mathrm{rcb}$ as a function of wavenumber, $k$, for mf-3au showing that the recombination time is at least three orders of magnitude smaller than the eddy turn-over times for all wavenumbers. The labels are the same as those used for Fig.\,\ref{fig:teddy-omega-mf}.}
\label{fig:teddy-mf}
\end{figure}

The recombination timescale divided by the eddy turnover time for mf-3au is shown in Fig.\,\ref{fig:teddy-mf}. Examining the timescales shows that $t_\mathrm{rcb} \ll t_\mathrm{eddy}$ in both the radial and the vertical direction and for all wavenumbers. This means that simulations performed using the single fluid approach should provide reliable results because chemical recombination takes place faster than the fluid moves material around thus ensuring that the ionisation fraction depends on the neutral density.

Despite the results presented here which suggest that $t_\mathrm{rcb} \ll 1/\Omega \ll t_\mathrm{eddy}$ it is an important detail to consider since the value of $t_\mathrm{rcb}$ is highly dependent on the chemical network used. The results will also be different once vertical stratification is considered.  

\section{Conclusions} 
\label{sec:conclusions}
We have performed two multifluid and two ideal global simulations to investigate the influence of non-ideal MHD effects. While local shearing box simulations provide vital clues as to the behaviour to be expected, in order to examine the interaction between adjacent regions governed by different physical processes and the global evolution of the PPD it is necessary to perform large-scale simulations.

For the multifluid simulations the weak Hall dominated region extends radially thus altering the dynamics of the system. Strong Maxwell stresses in the inner region of the disk give rise to angular momentum transport occuring via a flow exhibiting very low turbulence. These strong Maxwell stresses are accompanied by ordered magnetic structure for the multifluid simulations. The ideal MHD simulations have lower angular momentum transport but do display ordered magnetic structures. The Hall effect twists the magnetic field leading to the strong Maxwell stresses seen in the multifluid simulations. The transition from areas of structured magnetic field to more turbulent regions does not result in any global density structures, such as gaps, or in a sudden change in angular momentum transport.

We have introduced a dimensionless number, $\alpha_\mathrm{RM}$ (representing the ratio of the contributions to the $\alpha$ parameter due to Reynolds stresses and Maxwell stresses), to characterise accretion with $\alpha_\mathrm{RM} \gtrsim 0.1$ corresponding to turbulent transport. Determining the level of turbulence expected in PPDs is of importance for planet formation and the growth of planetesimals. It will also help distinguish between regions where MRI driven turbulence dominates and those regions launching MCWs.

In order to quantify the importance of turbulent mixing, we calculated the eddy turn-over time and compared this with an effective recombination timescale. We found that the recombination timescale is approximately three orders or magnitude smaller than the eddy turn-over time for all wavenumbers. This indicates that turbulent mixing is not significant. Nonetheless, these quantities should continue to be compared as chemical networks are updated and for vertically stratified simulations. Along with this we also find that the ionisation fraction of the disk is not seen to vary appreciably making these simulations comparable to single fluid non-ideal simulations.

\section*{Acknowledgements}
D.R.L. and T.P.D. would like to acknowledge PRACE for awarding them access to the resource JUQUEEN based in Germany at the J\"ulich Supercomputing Centre. D.R.L. and T.P.R. acknowledge support from Science Foundation Ireland under grant 13/ERC/I2907. D.R.L. would like to thank Antonella Natta and Aleks Scholz for numerous interesting discussions and for their contribution to improving the paper. The authors would like to thank the anonymous referee for many helpful comments.

\appendix
\section{Numerical tests}
Extensive numerical tests have been, and continue to be, performed on the HYDRA code. Tests of relevance to the simulations given in this work are presented in \citet{osullivan_2006,osullivan-2007} and \citet{downes-2014}. In particular, tests run on the ideal MHD setup \citep{downes-2014} include the Orszag-Tang vortex test (demonstrates the multi-dimensional performance of the code) and the Brio-Wu shock tube test (demonstrates the shock-capturing nature of the code). These tests give us some confidence that HYDRA performs well and is reliable within the general parameter ranges used in this work. 

\subsection{MRI growth rates}
To verify that the code HYDRA reproduces the expected MRI growth rates we compare the growth rates from linear analysis \citep{balbus_1991} with those calculated from the ideal MHD simulation, presented as res4-ideal, in \citet{okeeffe_2014}. This simulation is used, due to the data being outputted more frequently, instead of the simulations presented in this paper. From \citet{balbus_1991}, the most rapidly growing wavenumber has a growth rate of $\sim 0.75\Omega$, confirmed numerically by \citet{hawley_1991}, for instance. When considering the growth rates measured here it is essential to keep in mind that they are obtained from a global simulation. The growth rate is itself a local quantity and hence we must measure it at a particular radius. At $R = 2.2\,$au, where we determine the growth rates presented here, the system is perturbed both by the initial perturbation {\em and} by the growth of the instability at greater radii. Hence, we do not expect local growth rates measured from global simulations to be directly comparable with linear theory. 

\begin{figure}
\centering
 \includegraphics[width=0.5\textwidth]{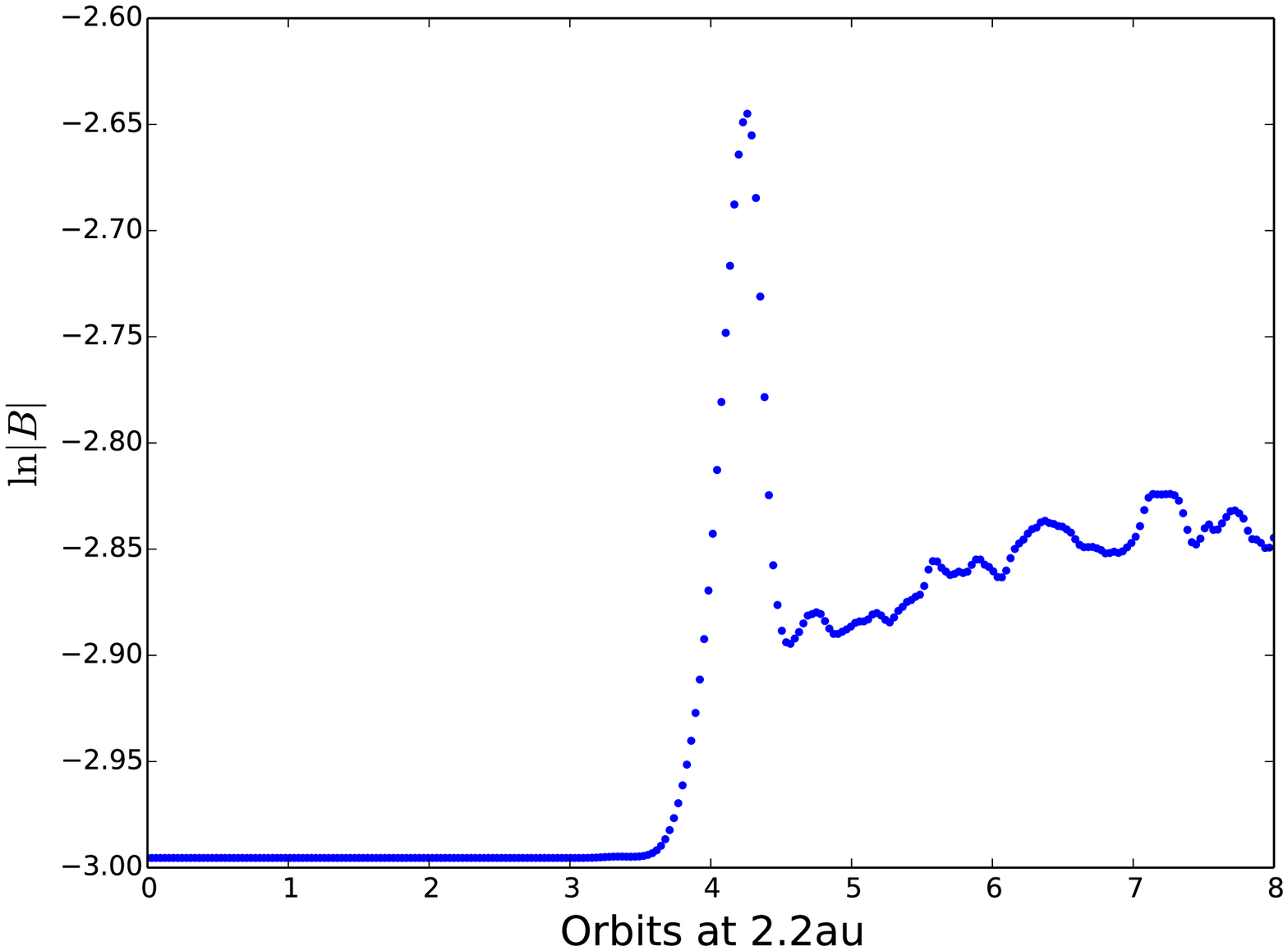}
  \caption{Log-linear plot of the magnitude of the magnetic field as a function of time, given in orbits calculated at 2.2\,au.}
\label{fig:gr}
\end{figure}

We plot the natural log of $|B|$ (averaged azimuthally) as a function of time in Fig.\,\ref{fig:gr} and measure the slope by a least squares fit to obtain the linear growth rate. At 2.2\,au the fastest growing wavelength is resolved by 7 gridzones and the growth rate is measured to be $\sim 0.62\Omega$. Bearing in mind the above considerations, this is broadly in line with the values found by \cite{balbus_1991} from linear analysis and the numerical simulations of \citet{hawley_1991}, verifying that HYDRA reproduces the MRI growth rate.

\begin{figure}
\centering
 \includegraphics[width=0.5\textwidth]{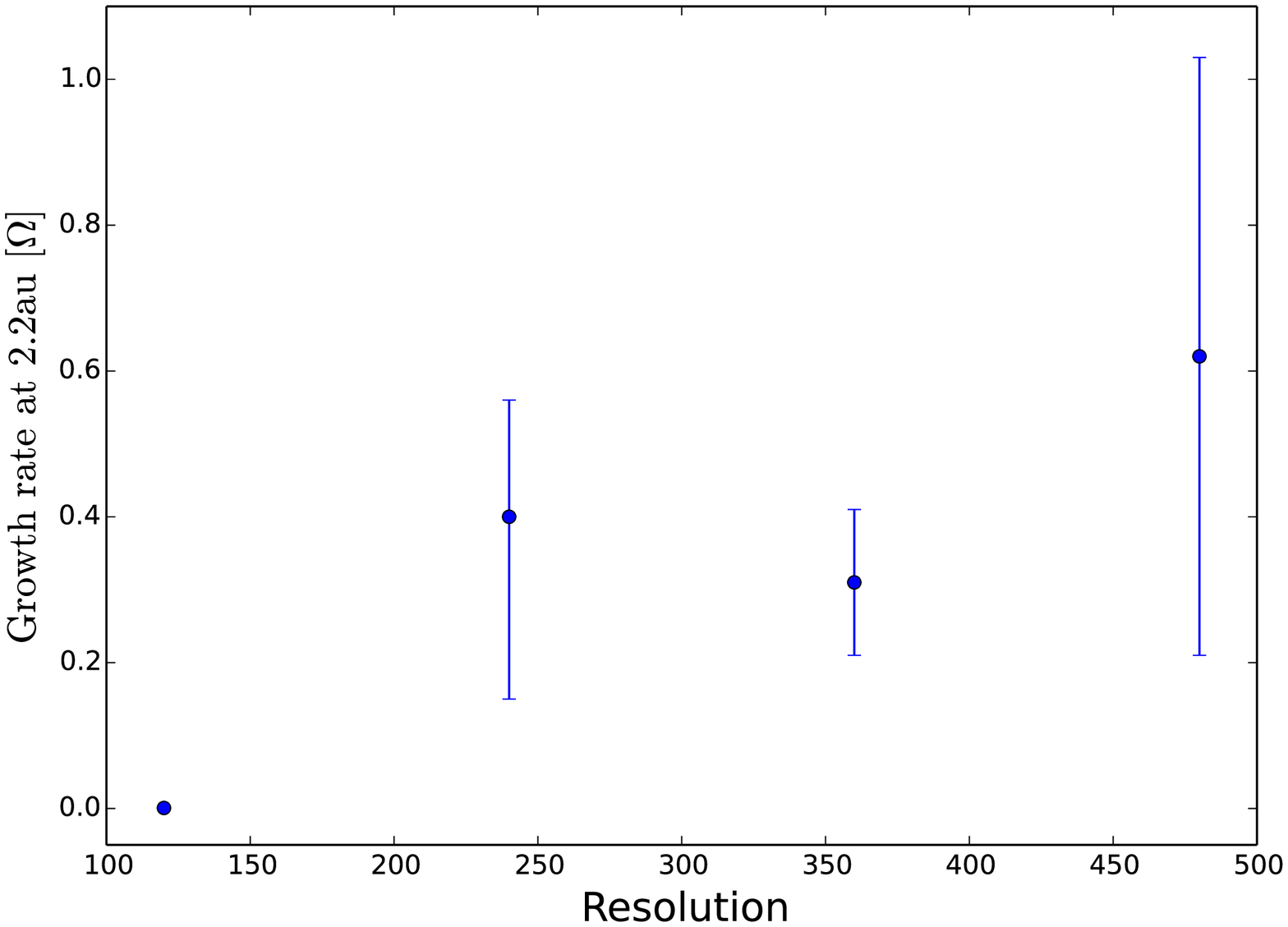}
  \caption{Plot of the growth rate of the MRI for ideal MHD as a function of resolution.}
\label{fig:gr-res}
\end{figure}

We also use the ideal MHD simulations (res1-ideal, res2-ideal, res3-ideal and res4-ideal) from \citet{okeeffe_2014} to plot the linear growth rate as a function of resolution, shown in Fig.\,\ref{fig:gr-res}. For each simulation the growth rate plotted is the average of four different different fits in time and the error bars given are the minimum and maximum value obtained for the linear growth rate. Considering the error bars the growth rate is converging as a function of resolution. In each case, the end time for measuring the linear growth rate was defined as the point at which the rate of growth began to decrease. The start time for measuring the growth rate was defined as 0.5 orbits before the end time. For res2-ideal, res3-ideal and res4-ideal the fits were performed between 5.33-5.82, 4.9-5.42 and 3.68-4.17 orbits (measured at 2.2\,au), respectively. These time intervals are not the same as the MRI begins to grow at later times for the lower resolution runs. For res1-ideal, the MRI only begins to grow at the very end of the simulation and the growth rate was measured between 196-244 orbits, corresponding to the leftmost data point in Fig.\,\ref{fig:gr-res}.
 
\newcommand\aj{AJ} 
\newcommand\actaa{AcA} 
\newcommand\araa{ARA\&A} 
\newcommand\apj{ApJ} 
\newcommand\apjl{ApJ} 
\newcommand\apjs{ApJS} 
\newcommand\aap{A\&A} 
\newcommand\aapr{A\&A~Rev.} 
\newcommand\aaps{A\&AS} 
\newcommand\mnras{MNRAS} 
\newcommand\pasa{PASA} 
\newcommand\pasp{PASP} 
\newcommand\pasj{PASJ} 
\newcommand\solphys{Sol.~Phys.} 
\newcommand\nat{Nature} 
\newcommand\bain{Bulletin of the Astronomical Institutes of the Netherlands}
\newcommand\memsai{Mem. Societa Astronomica Italiana}

\newcommand\apss{Ap\&SS} 
\newcommand\qjras{QJRAS} 

\bibliographystyle{mn2e}
\bibliography{../../donnabib}

\label{lastpage}

\end{document}